\documentclass[4pt,twocolumn]{article}
\pdfoutput=1
\usepackage[lmargin=0.75in,rmargin=0.75in,tmargin=1.0in,bmargin=1.0in]{geometry}
\usepackage{wrapfig}

\usepackage{graphicx}
\usepackage{pdfpages}

\usepackage[font={small}]{caption} 

\usepackage[square,numbers,sort]{natbib}
\usepackage{authblk}

\usepackage{hyperref} 

\begin{document}
\title{Zero-Shot Learning of Aerosol Optical Properties with Graph Neural Networks}
\date{}
\setcounter{Maxaffil}{1}

\author[a]{Kara D. Lamb}
\author[a]{Pierre Gentine} 

\affil[a]{Department of Earth and Environmental Engineering, Columbia University}

\maketitle

\begin{abstract} 
\textbf{Aerosols sourced from combustion such as black carbon (BC) are important short-lived climate forcers whose direct radiative forcing and atmospheric lifetime depend on their morphology. These aerosols' complex morphology makes modeling their optical properties difficult, contributing to uncertainty in both their direct and indirect climate effects. Accurate and fast calculations of BC optical properties are needed for remote sensing inversions and for radiative forcing calculations in atmospheric models, but current methods to accurately calculate the optical properties of these aerosols are computationally expensive and are compiled in extensive databases off-line to be used as a look-up table. Recent advances in machine learning approaches have shown the potential of graph neural networks (GNN's) for various physical science applications, demonstrating skill in generalizing beyond initial training data by learning internal properties and small-scale interactions defining the emergent behavior of the larger system. Here we demonstrate that a GNN trained to predict the optical properties of numerically-generated BC fractal aggregates can accurately generalize to arbitrarily shaped particles, even over much larger (10x) aggregates than in the training dataset, providing a fast and accurate method to calculate aerosol optical properties in models and for observational retrievals. This zero-shot learning approach could be integrated into atmospheric models or remote sensing inversions to predict the physical properties of realistically-shaped aerosol and cloud particles. In addition, GNN's can be used to gain physical intuition on the relationship between small-scale interactions (here of the spheres' positions and interactions) and large-scale properties (here of the radiative properties of aerosols).}
\end{abstract}

Carbonaceous aerosols such as black carbon (BC) are important short-lived climate forcers \cite[]{Bond2013, Liu2020}. To understand their impact on climate, accurate predictions of the optical properties of absorbing aerosols such as BC are needed in atmospheric models and observational retrievals: for estimating the top-of-the atmosphere radiative effects of black carbon \citep[]{Wu2016} and the impact of aged soot on cloud formation \cite[]{Lohmann2020}, for the calculation of the mass absorption coefficient of BC deposited on snow \cite[]{Schwarz2013}, for estimating the relative shortwave heating rates for different types of combustion aerosols \cite[]{Moteki2017}, for calculating particle-to-gas heat transfer to interpret laser-induced incandescence signals \cite[]{Michelsen2015}, for accurate inversions of imaging nephelometers \cite[]{Manfred2018}, for constraining the index of refraction of biomass burning aerosols \cite[]{Womack2020}, and for interpreting the optical properties of aerosols deposited on filters \citep[e.g.][]{Chakrabarty2011a,Chakrabarty2011b}. Accurate calculations of carbonaceous aerosol optical properties are also important for observational retrievals in other planetary atmospheres, as these aerosols may play a role in the radiative balance of e.g. the middle atmosphere of Jupiter \cite[]{Zhang2015}. 

\begin{figure}[h!]
\centering
 \includegraphics[width=0.8\linewidth]{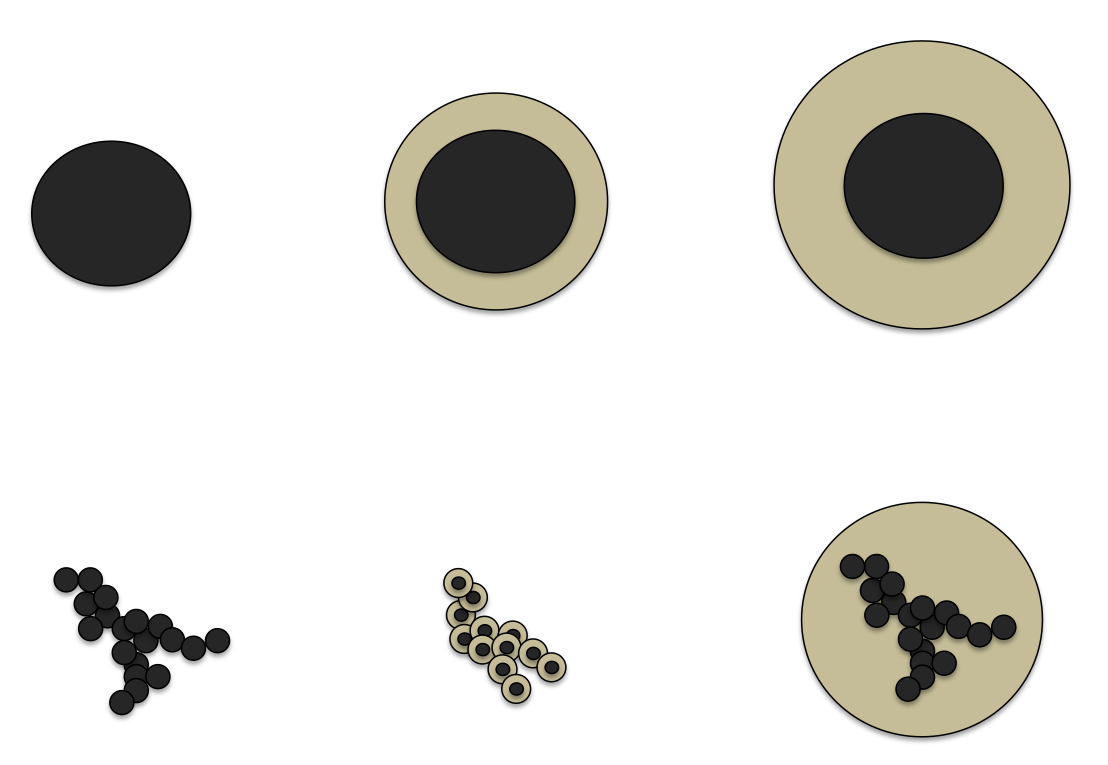}
\caption{\textbf{BC optical properties}. Top, from left to right: Equivalent volume sphere for bare BC, thinly coated BC, and thickly coated BC particles. Bottom, from left to right: Geometry of bare BC, thinly coated BC, and thickly coated BC as used in typical MSTM calculations.}
\label{bccartoon}
\end{figure}

BC particles in the atmosphere have a variety of sizes, shapes, and chemical compositions, all of which impact their optical properties (Figure \ref{bccartoon}). BC's optical properties depend on both the morphology of the primary (bare) BC particle, as well as its internal mixing with other materials (coatings) through the condensation of gas phase species during atmospheric aging. Both combustion conditions \cite[]{Wu2020} and atmospheric aging \cite[]{Wang2017} impact the morphology of these aerosols, which are fractal-like aggregates, typically embedded within (internally mixed) or attached to other aerosol components. The complex morphology of bare BC is generally not parameterized in models, although modeling bare BC as a sphere biases radiative forcing estimates, with too little warming by absorption and too much cooling by scattering \cite[]{Kahnert2020}. Internal mixing is modeled using a Mie Theory core-shell model, which approximates the bare BC portion as an absorbing "core", with a concentric sphere of "coating" material with an index of refraction characteristic of the internally mixed material. Several recent papers have demonstrated this Mie Theory core-shell approximation leads to an over-prediction of BC absorption in models by as much as a factor of 2 \cite[]{Fierce2020, Wu2020}. In addition, not only are more accurate calculations of BC optical properties needed to better constrain models to observations, but models need to be capable of representing the heterogeneity of optical properties in diverse aerosol populations \cite[]{Fierce2020, Wu2020}. 

While models and observational retrievals have generally relied on Mie Theory, more accurate methods to predict the optical properties for arbitrarily shaped particles such as the Multiple Sphere T-Matrix Method (MSTM) \cite[]{Mackowski1994,Mackowski1996}, the discrete dipole approximation (DDA) \cite[]{Purcell1973,Yurkin2007}, and the Generalized Multiple-Particle Mie (GMM) Theory \cite[]{Xu1995,Xu2001} have been developed. These methods approximate BC fractal aggregates as clusters of spheres (Figure \ref{bccartoon}) and provide exact analytical solutions to the time-harmonic Maxwell's equations for the multiple sphere system. However, these approaches are computationally expensive, often requiring hours or even days to compute the optical properties of single aerosol particles with complex morphologies \cite[]{Liu2019}. To mitigate this computational bottle-neck, pre-calculated databases of fractal aggregate optical properties using these exact analytical methods have recently been created \cite[]{Kahnert2010,Smith2014,Liu2019,Romshoo2021}, but such approaches are limited to linear interpolation within the data-bases' optical and morphological properties. There is still significant uncertainty about the fundamental properties of BC from different emission sources and under different combustion conditions, and the additional complexity of internal mixing with non-absorbing and absorbing materials during atmospheric aging \cite[]{Liu2020} would require these databases to cover a very large parameter space to accurately represent the range of conditions for BC aerosols observed in the atmosphere.  Moreover, observational inversions of BC have greater uncertainty when performed with only a subset of possible parameters.

Machine learning offers a promising approach for reducing computational bottle-necks by speeding up numerically-intensive aspects of atmospheric models \cite[]{Gentine2018,Rasp2018}. As such it could offer an efficient alternative approach to compiling pre-computed databases for BC's optical properties. However machine learning methods are traditionally strongly dependent on the data they are trained with, and struggle to generalize beyond the training distribution. An illustrative example previously investigated a machine learning approach to predicting BC's optical properties from its morphological parameters and index of refraction using a support vector machine (SVM) trained on accurate MSTM calculations but could not accurately predict the optical properties of aggregates with morphological parameters beyond those used in the initial training data set \cite[]{Luo2018}. Other brute force approaches such as neural networks (NN) or random forests (RF) will similarly struggle to generate realistic BC properties outside of the training datasets.

\begin{figure*}[h!]
\centering
 \includegraphics[width=1.0 \linewidth]{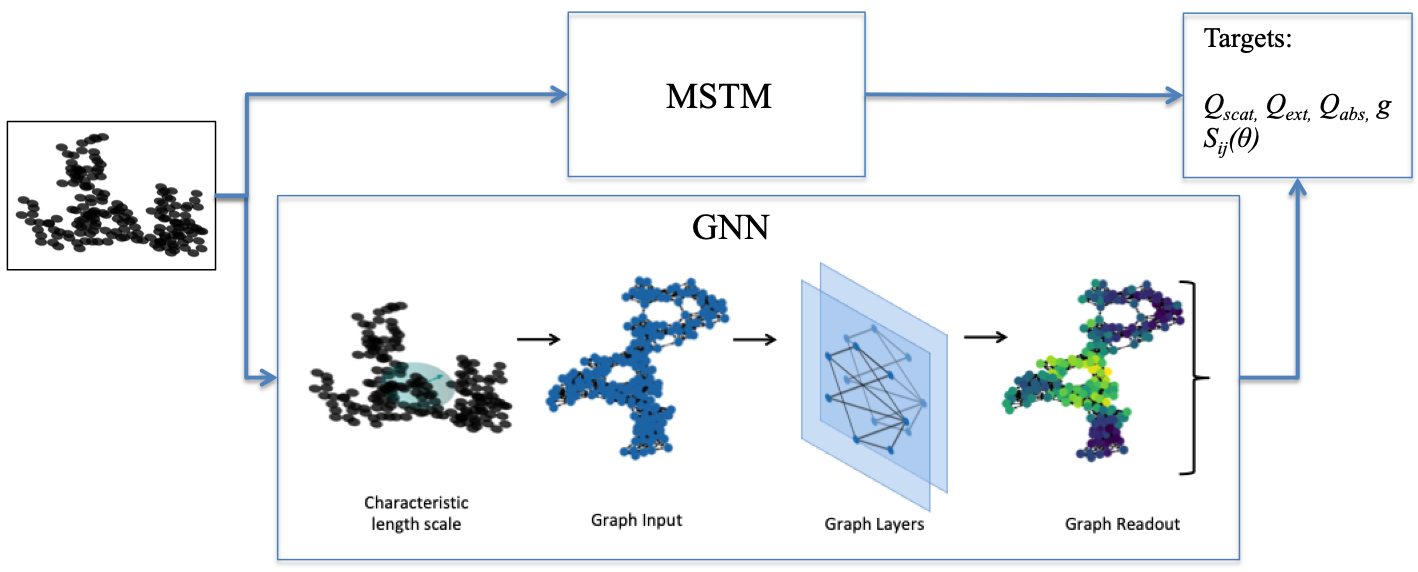}
\caption{A schematic of the GNN modeling approach for predicting aerosol optical properties. Accurate calculations of aerosol optical properties from the sphere positions are calculated using MSTM. For the GNN model, graphs are generated from aggregates by connecting spheres closer together than the characteristic length scale, $C$, of the aggregate (Eq. \ref{charlength}). Embeddings are learned for each node in the graph based on the central node features, the neighboring node features, and the edge features. These node-level embeddings are then aggregated together and a graph level prediction of the optical properties of the aggregate is made.\label{graphmodel}}
\end{figure*}

Here we show the optical properties of bare BC with complex morphology can be accurately predicted with a graph neural network (GNN) by representing BC fractal aggregates as networks of interacting spheres. GNN's are recently developed machine learning algorithms that learn on graph-structured data sets, allowing models to directly include arbitrary relational information \cite[]{Kipf2016,Battaglia2018}. These models have shown great promise in predicting the large-scale properties of structured physical science data-sets such as molecules \cite[]{Duvenaud2015,Gilmer2017}, protein-protein interaction networks \cite[]{Senior2020}, and glasses \cite[]{Bapst2020}. GNN's have demonstrated skill in predicting complex global features of physical systems through learning simpler local physics \cite[]{Xie2019}; here we demonstrate that through including local information about BC's structure, BC's global properties can be inferred. Importantly, because GNN's learn models for specific substructures (i.e. the nodes and their relationships with their neighbors in the graph), they are able to immediately generalize to graphs with arbitrary numbers of nodes; we exploit this feature of GNN's to predict the optical properties of BC aggregates that are significantly larger than those used in the training data set. This zero-shot learning (where models can immediately generalize to samples not represented in their original training data) paves the way towards new, flexible parameterizations of aerosol microphysical properties.

\section*{BC Fractal Aggregates as Networks}

\paragraph{Physical properties of bare BC}
Primary (bare) BC particles are fractal-like aggregates with geometries that can be described according to a statistical scaling rule as 
\begin{equation}
N_{s} = k_{f}\left( \frac{R_{g}}{a}\right) ^{D_{f}}
\end{equation}
where $a$ is the primary particle mean radius, $k_{f}$ is the fractal pre-factor, $D_{f}$ is the fractal (Hausdorff) dimension, $N_{s}$ is the number of primary spheres, or monomers, in the aggregate, and $R_{g}$ is the radius of gyration, defined as
\begin{equation}
R_{g}^{2} = \frac{1}{N_{s}}\sum_{n=1}^{N_{s}}(\mathbf{r}_{i}-\mathbf{r}_{0})^{2}
\end{equation} 
where $\mathbf{r}_{i}$ and $\mathbf{r}_{0}$ denote the $i$th monomer center and the center of mass of the cluster, respectively (assuming all monomers have the same mass \cite[]{Forrest1979}). 
In addition to the aggregate geometry, the basic physical properties of these particles follow this scaling law \cite[]{Filippov2000}. As a consequence of their fractal nature, aggregates are self-similar on different length scales. 

The fractal-like nature of these aerosols is a result of their formation from gas-phase precursors through the aggregation and growth of hydrocarbon clusters during incomplete combustion, although this process is not yet completely understood \cite[]{Johansson2018}. The initial morphology depends on both the combustion conditions and the emission source, with different observational methods also impacting the retrieved parameters \cite[]{Kahnert2020}. After their initial formation during combustion, atmospheric aging (due to cloud processing or the condensation of gas phase species) leads to these aerosols becoming more compact, causing $D_{f}$ to increase over time. This aging is expected to lead to a decrease in their top of the atmosphere radiative effects \cite[]{Wu2020}. Previous work has shown that $k_{f}$ determines the compactness of aggregate branches, although little is understood about $k_{f}$'s evolution over time \cite[]{Kahnert2020}.

\subsection*{Numerically-generated Fractal Aggregates}
To investigate how fractal aggregate particles can be modeled as networks of interacting spheres, we numerically generated fractal aggregates with $N_{s}$ spheres using a cluster-cluster algorithm \cite[]{Moteki2019} based on the one described in \cite[]{Filippov2000}, which uses a Monte Carlo approach to randomly generate aggregates with a specified fractal dimension $D_{f}$ and fractal pre-factor $k_{f}$. We generate Cartesian coordinates for the monomers in the aggregate in dimensionless coordinates by scaling by a factor of $k=\frac{2\pi}{\lambda}$, where $\lambda$ is the wavelength of the incident light.

\begin{figure}[ht!]
\centering
 \includegraphics[width=0.9\linewidth]{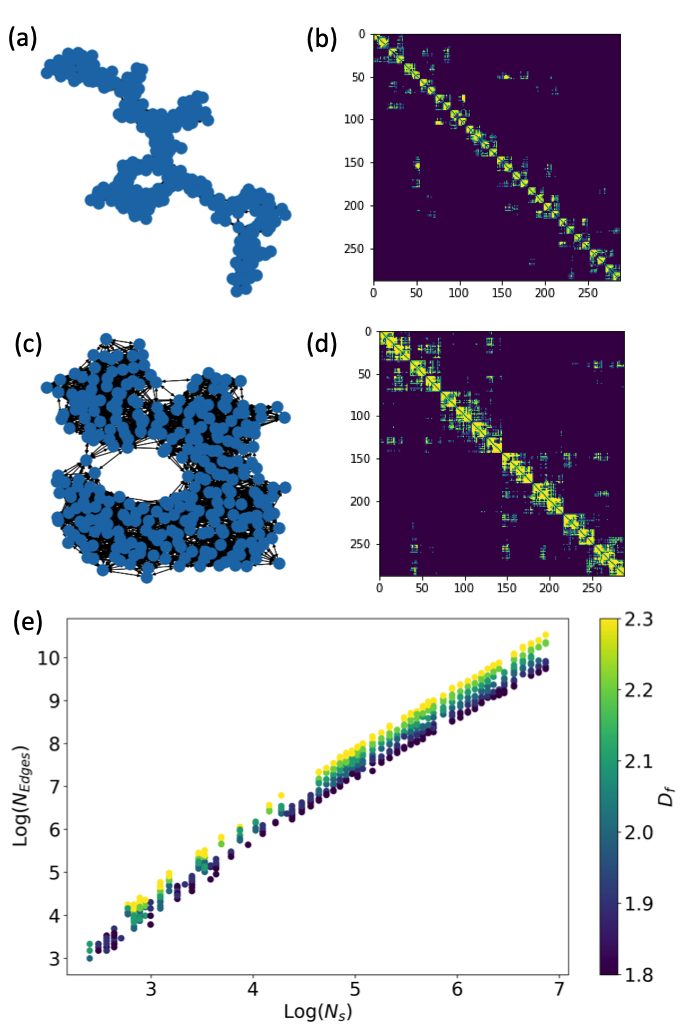}
\caption{Visualization of the graphs \textbf{(a,c)} and adjacency matrices \textbf{(b,d)} for fractal aggregates with the same number of spheres ($N_{s}$=288) but different fractal dimensions. $D_{f}$=1.8 for (a,b) and $D_{f}$=2.3 for (c,d). \textbf{(e)} The number of edges scales with the total number of spheres in the aggregates ($N_{s}$) and the fractal dimension of the aggregates ($D_{f}$).\label{graphaggs}}
\end{figure}

\subsection*{Characteristic length scale}
To represent fractal aggregates as graphs, monomers with center positions closer together than the characteristic length scale $C$ of the network, 
\begin{equation}
C=X_{v}Log(N_{s}) 
\label{charlength}
\end{equation}
are connected, where $X_{v}=ka$ is the monomer size parameter (Fig. \ref{graphmodel}). This assumption derives from the characteristic length scale of a network with $N_{s}$ nodes \cite[]{Albert2002}. Multiplying by $X_{v}$ gives a consistent number of edges irregardless of the size parameter of the aggregate, such that aggregates with the same fractal parameters but different size parameters would be encoded within the same graph structure. An example of the resulting undirected graph structure and adjacency matrix for two different aggregates with different fractal dimensions but the same number of monomers is shown in Figure \ref{graphaggs}a-d. This scaling encodes the density of edges in local neighborhoods relative to the fractal dimension of the aggregate, irregardless of the actual size of the aggregate. The total number of edges in the graph is then proportional to both $N_{s}$ and $D_{f}$ (Figure \ref{graphaggs}e), with the average degree of nodes increasing relative to $D_{f}$ (SI Fig. 4a). The degree distribution of nodes also depends on the fractal pre-prefactor $k_{f}$(SI Fig. 4b).

\section*{GNN model for BC optical properties}

Accurate solutions for the electromagnetic scattering and absorption properties for multiple sphere clusters (as BC aggregates are typically modeled) is computationally expensive because a full-wave optics treatment is needed. In the general case, spheres interact with one another, and the total scattering field component is a superposition of the components radiated from each sphere in the system \cite[]{Bohren2008}. While the solution for the continuity equation at the surface of each sphere in the system can be solved analytically by expanding the incident and scattered fields from each sphere in terms of vector spherical wave functions, this approach generates a very large system of coupled linear equations that must be solved iteratively \cite[]{Mackowski2011}. Additional details about the formal solution are given in Supplementary Information.

While this approach provides a fully analytical solution for light scattering from the multiple sphere cluster, the computational time for these brute-force approaches scale significantly with $N_{s}$ and $X_{v}$ as they do not take into account specific details of BC's topological structure, which could lend itself to model order reduction. Filippov et al. \cite[]{Filippov2000} previously explored the relationship between the morphology of BC and their aggregate physical properties using the Rayleigh-Debye-Gans (RDG) approximation and found that aggregates with similar fractal parameters also have similar physical properties. Recent work in \cite[]{Liu2019} found empirical relationships between the optical properties of aggregates and their morphological parameters using extensive MSTM calculations. Machine learning offers an alternative approach for learning relevant predictors without the need for human-defined features; GNN's in particular can learn features that correspond to the relationship between the nodes (the individual spheres) and the large scale physical properties of the aggregates.

To investigate the connection between BC's fractal structure and its optical properties, we trained a GNN to predict the optical properties of BC aggregates, using the values from an analytical solution for the electromagnetic scattering and absorption properties from MSTM as ground-truth (Figure \ref{graphmodel}). We tested several different GNN approaches (See SI) and found an Interaction network (IN) \cite[]{Battaglia2018} gave the best performance for predicting both the integral and angle-resolved optical properties. The IN (Fig. \ref{graphmodel}) is based on message passing, where nodes send and receive messages along edges from their neighbors. The messages are aggregated for each node and the nodes are updated based on the central node features and the messages received from neighboring nodes. Graph level predictions are made by aggregating the updated node embeddings from all the nodes in the graph and then applying a graph level model to the aggregated node embeddings (the readout in Figure \ref{graphmodel}). Here we predict the total extinction, scattering, and absorption efficiencies $\langle Q_{ext}\rangle $, $\langle Q_{scat}\rangle $, $\langle Q_{abs}\rangle $, the asymmetry parameter, $g$, and the angle-resolved elements of the scattering phase matrix $S_{ij}(\theta)$ for the orientation-averaged case. (See Methods for discussion of aerosol optical properties and data sets).

\begin{figure}[ht!]
\centering
\includegraphics[width=0.95\linewidth]{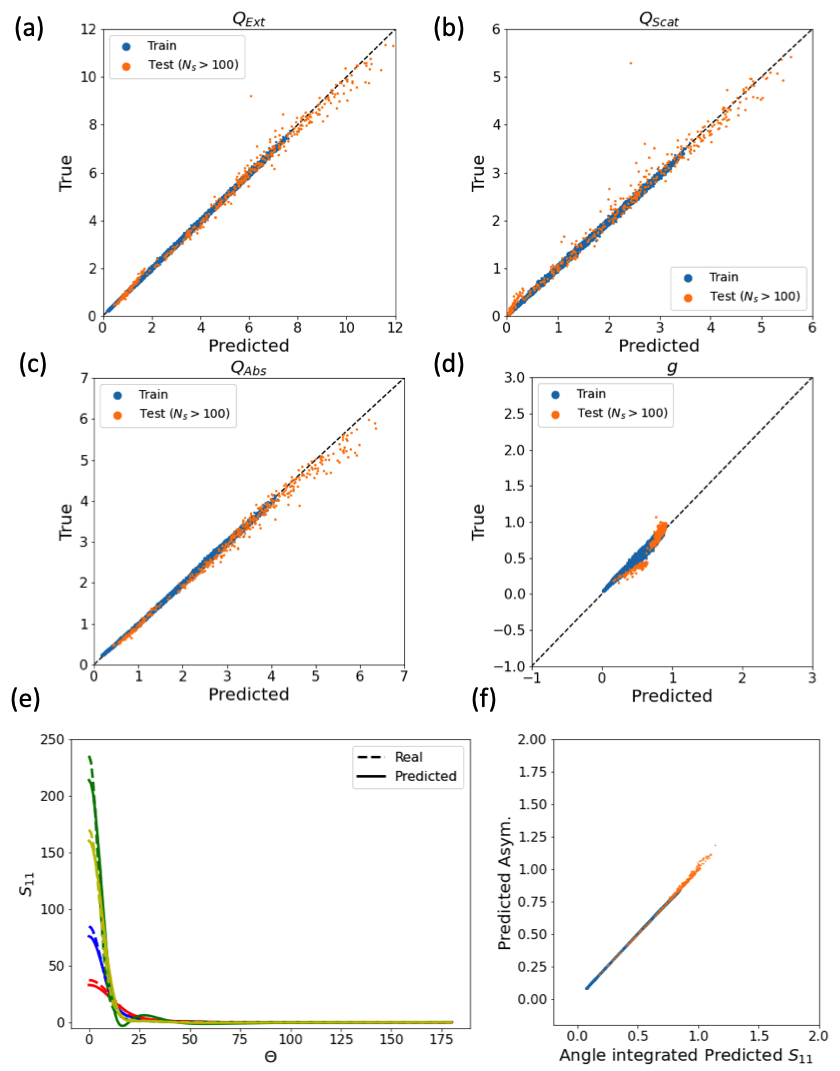}
\caption{\textbf{(a-d)} The true vs. predicted values for the efficiencies and the asymmetry parameter for the training ($N_{s}<100$) and zero-shot test ($N_{s}=100-1000$) data sets are shown. \textbf{(e)} Comparison of $S_{11}(\Theta)$ predictions for 4 randomly chosen aggregates in the test set: Blue ($N_{s}=128$,$X_{v}=0.9$,$n_{k}=1.6+i0.6$),
Red ($N_{s}=640$,$X_{v}=0.3$,$n_{k}=2.0+i1.0$),
Green ($N_{s}=960$,$X_{v}=0.7$,$n_{k}=1.8+i0.8$), Yellow ($N_{s}=416$,$X_{v}=0.9$,$n_{k}=2.0+i1.0$). \textbf{(f)} The predicted $S_{11}$ integrated over the solid angle (Eq. \ref{S11constraint}) vs. the predicted value for asymmetry parameter ($g$) for all of the aggregates in the training and test sets. \label{results}}
\end{figure}

For each training example, we input $X_{v}$, the real part of the index of refraction $Re(n_{k})$ (since we consider only cases where the imaginary part is $Im(n_{k})=1-Re(n_{k})$), and the dimensionless coordinates of each sphere as node features. As edge features, we use the distance between neighboring spheres. We trained the model using 15314 aggregates from the training data set, as the training loss did not significantly decrease with additional samples (SI Figure S13); training data sets as small as 3000 aggregates showed reasonable generalization performance. The training data set consisted of aggregates with a small number of monomers ($N_{s} < 100$). We tested the model on an independent test set of 7656 aggregates with the same distribution of parameters as the training data set ($N_{s} < 100$). We further investigated the generalizability of the model on an independent zero-shot test set of 440 aggregates that were significantly larger  $(100 < N_{s} < 1000)$ than the ones the model was trained on. An additional 440 large aggregates were used as a zero-shot validation data set to determine which model architecture provided the best zero-shot performance (SI Figures S9-S12). $N_{s}=100$ was chosen as the maximum size for aggregates in the training data set as smaller maximum sizes increased the bias in the zero-shot performance (SI Figure S14). The zero-shot test data set was evenly distributed among the aggregate parameters (Figure S2) to provide an estimate of generalization performance across the full parameter space. 

Figure \ref{results}a-d shows the IN predictions compared to the actual values for $\langle Q_{ext}\rangle $, $\langle Q_{scat}\rangle$, $\langle Q_{abs}\rangle$, and $g$ with the training data shown in blue and the zero-shot test data sets shown in orange, while Figure \ref{results}e shows the predictions for the $S_{11}(\theta)$ element of the scattering phase matrix for several different aggregates in the test sets. Predictions for both integral and angle resolved optical properties were reasonable across the entire range of size parameters ($X_{v}$=0.1 to 1.0), indices of refraction $n_{k} = 1.4+0.4i$ to $n_{k}=2.0+1.0i$, and fractal parameters, with predictions for the integral optical properties in the test data set within 10\% of the true values. For $S_{11}$, both the magnitude and functional form were well-approximated across the range of parameters in the test set, although the model did deviate slightly more from the true values for larger $N_{s}$ and $X_{v}$ (e.g. the green line in Fig. \ref{results}e). Predictions for the entire angle-resolved scattering phase matrix elements $S_{ij}(\theta)$, for $j \geq i$, were also reasonable (See SI Figure S16). For the test data with the same distribution of parameters as the training data set ($N_{s} < 100$), the model predictions were very close to the true values; these results are shown in SI Figure S17. 

In addition to generalizability, the IN model demonstrated physical consistency in its predictions for the aggregate optical properties. The 3 scattering efficiencies are not independent, as $\langle Q_{sca} \rangle+\langle Q_{abs} \rangle=\langle Q_{ext} \rangle$. The model directly inferred this dependency for both the training and test sets without imposing this as a constraint. Additionally, integrating $S_{11}$ over the solid angle is equivalent to $g$ \cite[]{Bohren2008}, 

\begin{equation}
g = \frac{1}{2}\int S_{11}(\theta)cos(\theta) d\Omega = \frac{1}{2}\int_{0}^{\pi} S_{11}(\theta)cos(\theta) sin(\theta) d\theta
\label{S11constraint}
\end{equation}
Without explicitly imposing this integral constraint, the model predictions were consistent with this constraint (Figure \ref{results}f). 

\subsection*{Analysis of the GNN predictions}
To understand how the IN model predicts the optical properties of BC fractal aggregates, including those much larger than the model was trained on, we emphasize that the graph input for the model does not directly include $D_{f}$ or $k_{f}$ as features but rather the fractal structure is implicitly encoded as the interactions between the neighboring spheres. The previously used SVM approach to predict BC's optical properties included $N_{s}$, $D_{f}$, and $k_{f}$ as features to predict $\langle Q_{ext}\rangle $, $\langle Q_{scat}\rangle$, $\langle Q_{abs}\rangle$, and $g$ \cite[]{Luo2018}. Since the network structure in the IN approach is directly learned from the sphere positions, and the model is learned at the node level, the IN approach can generalize beyond its initial training set for these morphological parameters to unseen configurations.

The generalization of the IN model to a range of $D_{f}$ is an important feature, as it is challenging to find approximations that are valid across fractal dimension \cite[]{Sorensen2001}. Because the IN model learns about the local neighborhood of each sphere, it is able to more accurately estimate the impacts of screening on absorption and scattering than the RDG approximation, \cite[]{Sorensen2001,Bohren2008}, an approach often used to approximate the optical properties of BC aggregates in a computationally efficient manner as an improvement on the equivalent sphere model. RDG assumes that individual monomers only interact with the incident electromagnetic field (neglecting multiple interactions), which can lead to absorption being under-predicted by 10-20\%, and significantly under-predicting $g$ by more than a factor of 10 \cite[]{Kahnert2020}. 
The IN model effectively learns, in an unsupervised manner, a simplified sphere level model that more fully captures the complexity of the optical properties of the full analytical solution \cite[]{Kahnert2020}. 

The optical properties of aggregates in this regime can be modeled with the assumption of a fairly shallow graph model (for the IN model a single layer performed best; for the GCN little improvement was seen beyond 3 or 4 layers, Figure S5), suggesting that the majority of the structure influencing the optical properties of aerosols in this regime can be approximated from local interactions. We also investigated using a length scale of $C=X_{v}Log(N_{s})/Log(Log(N_{s}))$ (characteristic of scale-free networks) to form graphs from aggregates \cite[]{Albert2002}, rather than Eq. \ref{charlength}. This length scale has the advantage that the degree of each node scales less quickly with $N_{s}$, but the IN model performed worse in this case. This indicates that including a larger local neighborhood at each layer (Eq. \ref{charlength}) is more informative for the model.

\section*{Discussion and Outlook}
The network approach presented here provides a new framework for understanding the microphysical relationship between the morphological properties of BC and its larger scale physical properties. Here we have chosen to focus on the prediction of optical properties for numerically generated fractal aggregates, as the generation of these aggregates from combustion processes and their transformation during atmospheric aging is not yet completely understood. However, applying network theory to atmospheric aerosols suggests new directions for thinking about the generation of these fractal aggregates through combustion processes due to the connection between complex networks and percolation theory \cite[]{Deprez2015}. Here we have used a cluster-cluster algorithm, although previous work has noted that the morphology of numerically generated fractal aggregates depends not only on the parameters ($N_{s}$, a, $D_{f}$, and $k_{f}$) defining the shape of the aggregate, but also on which algorithm is used to generate the sphere positions (e.g. diffusion-limited aggregation or diffusion-limited cluster aggregation) \cite[]{Sorensen1997,Filippov2000}. The network approach provides a new framework from which to understand how realistically numerical algorithms reproduce the properties of aerosols formed during incomplete combustion through comparison of their network characteristics \cite[]{Albert2002}. This approach may also be useful for inferring 3 dimensional structure of aggregates from 2 dimensional transmission electron microscope (TEM) images of these aerosols \cite[]{Chakrabarty2011a,Chakrabarty2011b}, since it relates the relative positions of spheres to their overall morphological features; 2D methods have previously been shown to systematically underestimate the fractal dimension of BC \cite[]{Adachi2007}. 

As a proof of concept we have trained a GNN to predict the optical properties of bare BC fractal aggregates with a range of different fractal parameters. This study demonstrates that modeling aerosol fractal aggregates as networks of interacting spheres provides morphological information that allows the machine learning model to extrapolate far beyond their initial training data set. This approach may also be useful for other fractal systems found in nature, such as turbulence, vegetation, or river networks.

BC in the atmosphere is typically internally mixed. The GNN approach provides an obvious extension to internally mixed aerosols (Figure \ref{bccartoon}), as the thickness of coatings and their indices of refraction or organic fraction could be included as additional node-level features (in the thinly coated case) or graph-level features (for the thickly coated case). Other factors influencing the optical properties of aggregates such as "necking" between overlapping monomers could be included as edge features. Because atmospheric aerosol retrievals rely on orientation averaged parameters, models for predicting the scattering phase function should be equivariant under rotations. Recently developed equivariant machine learning methods \cite[]{Thomas2018,Kondor2018,Miller2020,Satorras2021} may provide improved prediction of the orientation averaged optical properties.

Uncertainty in BC direct radiative climate effects is attributable to multiple factors, including BC's emissions, lifetime, atmospheric processing, and optical properties \cite[]{Bond2013,Wang2016,Liu2020}; the GNN approach could help resolve this uncertainty by improving both the interpretation of BC observations and by allowing BC's morphology to be accurately represented in atmospheric models in a computationally efficient manner. As a greater understanding of BC's physical properties from different source contributions and atmospheric aging pathways becomes available through laboratory and observational studies \cite[]{Wang2017,Fierce2020,Wu2020}, the major remaining hurdle to accurately representing BC in models will be computational. 

While previous exact analytical methods have computational wall-times scaling from hours to days for larger aggregates, inference is on the order of $<$ 0.3 seconds per aggregate for the trained GNN model (On a CPU-- see SI Figure S15). The computational time for these exact analytical methods has precluded exact calculations of aerosol optical properties being used in models or observational retrievals. CELES, a CUDA-accelerated version of MSTM capable of running on a GPU, demonstrated a factor of 1.5-6 times speed up over MSTM, but was still too slow to be implemented online in models \cite[]{Egel2017}. The significantly faster time-scale for the GNN model, as well as its generalizability to arbitrarily shaped aggregates compared to more standard ML methods, has the potential to transform existing model parameterizations for BC. For MSTM computational wall times scale with $N_{s}$, $X_{v}$, and $D_{f}$; while the total inference and memory scales with $N_{s}$ and $D_{f}$ in the GNN approach, it is no longer a function of $X_{v}$. 

We have focused here on the forward problem of predicting the optical properties of BC given an assumed single particle morphology; however such an approach may also be useful for the inverse problem, i.e. inferring the morphology given the scattering phase function and integral optical properties. This approach could also provide insight into other physical properties which require detailed information about particle morphology \cite[]{Filippov2000}, such as energy and heat transfer between aggregates and the surrounding gas needed to develop physical models of laser-induced incandescence \cite[]{Bambha2015, Michelsen2015}. Radiative transfer calculations for mineral dust and ice crystals also rely on detailed information about particle morphology, suggesting that the GNN approach would be useful for modeling their optical properties as well. This approach could mitigate several long-standing issues with model parameterizations and observational retrievals for these species, by providing flexible parameterization of arbitrarily shaped aerosol and cloud particles that are fast enough to be deployed online in atmospheric models. 

Finally, these methods have potential for new applications of machine-learning assisted materials discovery \cite[]{Moosavi2020,Mirhoseini2021}. Proposed geo-engineering approaches to mitigate global or regional impacts of climate change, such as stratospheric aerosol injection, marine cloud brightening, or precipitation enhancement, rely on the development of novel aerosol materials. Generative graph models could be used to determine optimal aerosol morphologies resulting in physical properties specific to these applications at a fraction of the cost of traditional numerical methods \cite[]{De2018}. 

\small
\section*{Methods}
\subsection*{Numerical aggregate properties}
Primary clusters of size $N_{c}=3,4,5,7,9,11,13,15,17,19$ were used to generate aggregates between $N_{s}$=8 to 960 spheres with fractal dimensions between $D_{f}$=1.8 to 2.3. Following \cite[]{Liu2019}, we assume a fractal pre-factor of $k_{f}$=1.2 (for the aggregates used in the MSTM calculations). We also investigated the network parameters of aggregates with $k_{f}=1.0-1.5$, for a given $D_{f}=1.8$ (Figure S4b). Aerosols are assumed to consist of isotropic, homogeneous spheres, with size parameters $X_{v} =$ 0.1, 0.3, 0.5, 0.7, 0.9, and 1.0, corresponding to monomer radii between 7 to 72 nm for incident light at 450 nm and 10 to 104 nm at 650 nm. For each primary cluster size and fractal dimension, 10 aggregate realizations were randomly generated. 
\subsection*{Aerosol optical properties}
For radiative transfer applications, the orientation-averaged total scattering $\langle Q_{sca} \rangle$, extinction $\langle Q_{ext}\rangle$, and absorption efficiencies $\langle Q_{abs}\rangle$, as well as the asymmetry parameter $g = \langle C_{sca} cos(\theta')\rangle$ are typical parameters that are needed ($C_{sca}$ is the scattering cross-section which is related to the efficiency as $Q_{sca}=C_{sca}/(pi*a_{agg}^{2})$, where $a_{agg}$ is the effective radius of the aggregate). The asymmetry parameter relates the amount of forward to back-scattered light. Other parameters relevant for radiative transfer, such as the single scattering albedo (SSA), can be derived from these parameters (SSA = $Q_{sca}/Q_{ext}$). The mass absorption coefficient (MAC) or mass extinction coefficient (MEC) are typically used to relate emissions of these aerosols to their direct radiative effects, and they are sometimes estimated theoretically from $\langle C_{abs}\rangle$ or $\langle C_{ext}\rangle$ with assumptions about particle density. 

The scattering phase function relates the incident and scattered Stokes parameters, e.g. it indicates how light scattering from the particle is transformed relative to incident light in terms of its intensity and polarization state \cite[]{Bohren2008}. Here we assume initially unpolarized incident light, in which case the $S_{11}$ element specifies the angular distribution of the intensity of scattered relative to incident light. The scattered light is partially polarized, with degree of polarization given by $\sqrt{(S_{21}^{2}+S_{31}^{2}+S_{41}^{2})/S_{11}^{2}}$.

\subsection*{MSTM calculations of bare BC optical properties}
 To determine the ground-truth optical properties for the BC fractal aggregates generated by the cluster-cluster algorithm we use the Fortran-90 implementation of the multiple-sphere T-matrix code as described in \cite[]{Mackowski2011}, which can run on a high-performance, parallel based computational platform. This code numerically solves for electromagnetic wave scattering from multiple (non-overlapping) sphere systems for either a fixed or random (orientation-averaged) orientation with respect to an incident plane wave. 
Here we have focused on calculation of random orientation optical properties, which utilizes the T-matrix procedure developed in \cite[]{Mackowski1996}. We assume indices of refraction consistent with a range of values from the literature for BC at 550 nm: (1.4+0.4\textit{i}, 1.6+0.6\textit{i}, 1.8+0.8\textit{i}, 2.0+1.0\textit{i}). MSTM calculations were performed for these range of indices of refraction for 57,556 numerically generated aggregates for $N_{s} < 100$; we used randomly chosen aggregates from this data set for the training, validation,  and test sets for the model. To test the zero-shot performance, MSTM calculations were performed for 880 aggregates with these parameters in the size range $100 < N_{s} < 1000$; we randomly split this data into a zero-shot validation data set to evaluate the model's performance and an independent zero-shot test data set. A summary of the range of parameters for each data set is given in Table S1. The distribution of parameters among the small ($N_{s}<100$) and large ($N_{s}>100$) aggregates are shown in Figures S1 and S2, and the integral optical properties calculated with MSTM are shown in Figure S3.

\subsection*{Graph Neural Networks}
We used Pytorch Geometric \cite[]{Fey2019} to implement the GNN models. Several GNN approaches were tested, including a simple graph convolutional network (SGC) \cite[]{Wu2019}, a graph convolutional network (GCN) \cite[]{Kipf2016}, and an interaction network (IN) \cite[]{Battaglia2018}  (See Supplementary Information for additional details of the graph models and a comparison of performance metrics among different model parameters and targets). The best performance for the integral optical properties used an IN model with a hidden layer size of 300 for both the node and edge models, and a message size of 100. Both the node and edge models are MLPs with ReLU as non-linear activation function between layers. Aggregation for the edge model is addition, with global mean pooling followed by dropout (p=0.5) and a linear layer of size 100 as the global aggregation function. For the prediction of $S_{11}$ we found that adding a fully connected node to each graph slightly improved the zero-shot performance. The model architecture was the same as that used to predict the integral optical properties. A batch size of 20 was used (training with a batch size of 2 led to slower training but did not lead to significantly worse performance). For the graph regression task, MSE loss was assumed. We trained the GNN models on a Nvidia RTX 8000 GPU.

\onecolumn

\clearpage
\includepdf[pages=1-last]{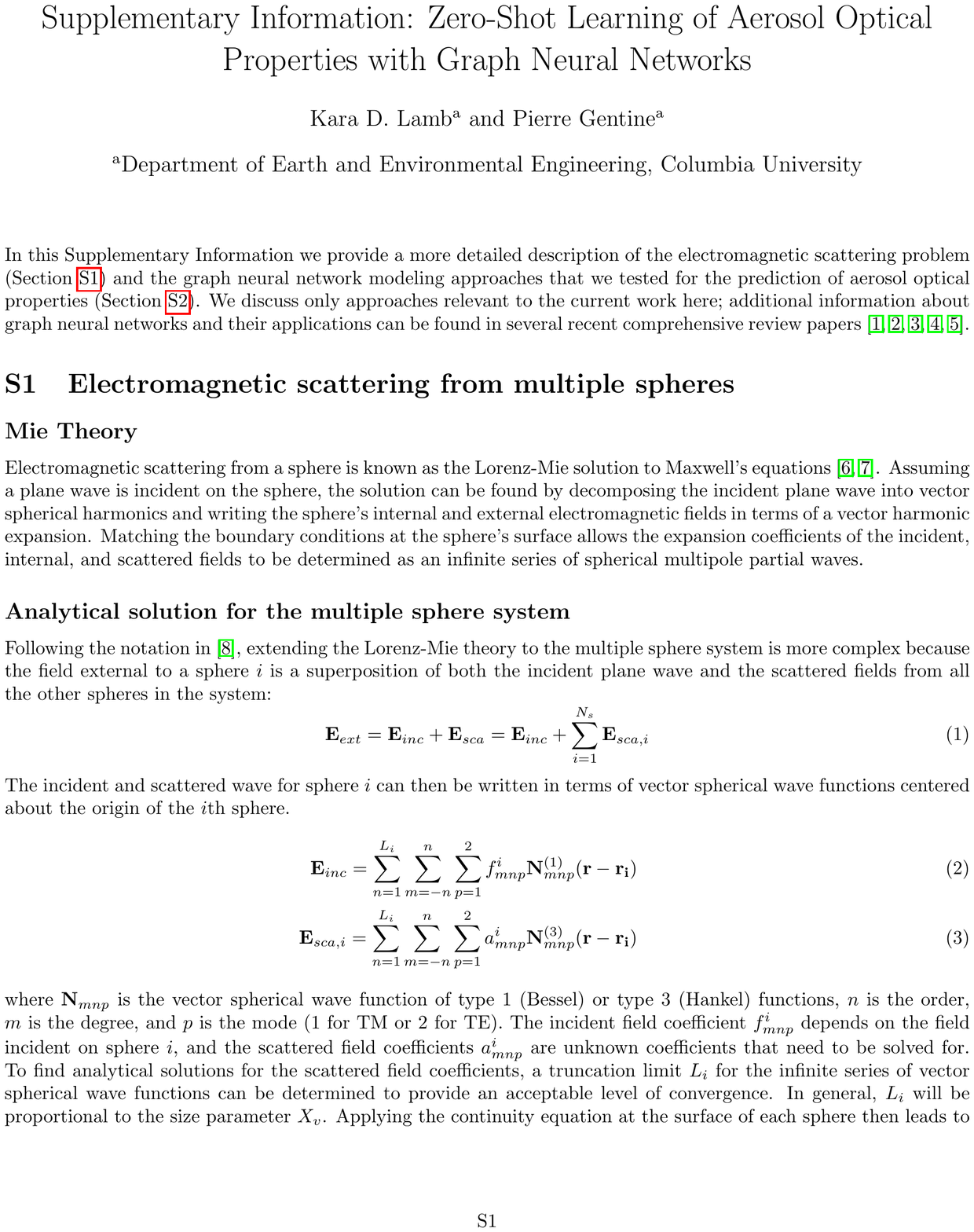}

\end{document}